\documentclass[10pt,  twocolumn, twoside, letterpaper, conference, final]{IEEEtran}
\IEEEoverridecommandlockouts

\usepackage{cite}
\usepackage{amsmath,amssymb,amsfonts}
\usepackage{algorithmic}
\usepackage{graphicx}
\usepackage{subcaption}
\usepackage{textcomp}
\usepackage{url}
\usepackage{array}
\usepackage{stfloats}
\usepackage{cite}
\usepackage{mathtools}

\usepackage{graphicx}
\usepackage{multirow}
\usepackage{textcomp}
\usepackage{diagbox}
\usepackage{colortbl}
\usepackage{xcolor}
\def\BibTeX{{\rm B\kern-.05em{\sc i\kern-.025em b}\kern-.08em
    T\kern-.1667em\lower.7ex\hbox{E}\kern-.125emX}}
\begin{document}

\title{A Pragmatical Approach to Anomaly Detection Evaluation in Edge Cloud Systems}

\author{\IEEEauthorblockN{Sotiris Skaperas\IEEEauthorrefmark{1}\IEEEauthorrefmark{2}, 
Georgios Koukis\IEEEauthorrefmark{3}\IEEEauthorrefmark{2},
Ioanna Angeliki Kapetanidou\IEEEauthorrefmark{3}\IEEEauthorrefmark{2},
Vassilis Tsaoussidis\IEEEauthorrefmark{3}\IEEEauthorrefmark{2},
Lefteris Mamatas\IEEEauthorrefmark{1}\IEEEauthorrefmark{2}}
\IEEEauthorblockA{\IEEEauthorrefmark{1} Department of Applied Informatics, University of Macedonia, Thessaloniki Greece} 
\IEEEauthorblockA{\IEEEauthorrefmark{2} Athena Research and Innovation Center, Greece}
\IEEEauthorblockA{\IEEEauthorrefmark{3} Department of Electrical and Computer Engineering, Democritus University of Thrace, Greece}

Emails: \{sotskap, emamatas\}@uom.edu.gr, \{gkoukis, ikapetan, vtsaousi\}@ee.duth.gr
}

\maketitle
\begin{abstract}
Anomaly detection (AD) has been recently employed in the context of edge cloud computing, e.g., for intrusion detection and identification of performance issues. However, state-of-the-art anomaly detection procedures do not systematically consider restrictions and performance requirements inherent to the edge, such as system responsiveness and resource consumption. In this paper, we attempt to investigate the performance of change-point based detectors, i.e., a class of lightweight and accurate AD methods, in relation to the requirements of edge cloud systems. Firstly, we review the theoretical properties of two major categories of change point approaches, i.e., Bayesian and cumulative sum (CUSUM), also discussing their suitability for edge systems. Secondly, we introduce a novel experimental methodology and apply it over two distinct edge cloud test-beds to evaluate the performance of such mechanisms in real-world edge environments. Our experimental results reveal important insights and trade-offs for the applicability and the online performance of the selected change point detectors. 
\end{abstract}


\section{Introduction}
Edge networks have emerged as an ideal paradigm to support services that require minimal latency and high bandwidth, including the ever-growing Internet-of-Things (IoT) applications \cite{yu2022edge}. However, they often exhibit inherent resource constraints and resource optimization is essential to reap the benefits of the operation of machine learning (ML) solutions.

Anomaly detection (AD) is a broad category of ML techniques aiming on identifying patterns that significantly deviate from the ``normal" (i.e., expected) behavior, currently used in various network-related problems, e.g., intrusion detection \cite{aldweesh2020deep}. In this context, change point (CP) analysis \cite{poor2009quickest} is a suitable candidate for AD in resource constrained environments, since it is aligned to the ”quickest” detection of an abrupt change, also ensuring low computational complexity. Moreover, stemming from the class of sequential statistical analysis \cite{polunchenko2012state}, these approaches bypass the need for exhaustive feature mining and labeling to achieve adequate performance.       

This framework has been efficiently incorporated in the cloud or wireless side of the edge, for instance we utilize them: (i) to trigger elasticity in unikernel-based edge clouds \cite{valsamas2018elastic}; (ii) to detect denial of service attacks in software-defined wireless sensors networks \cite{segura2020denial}; and (iii) as a link reliability detector in software-defined smart-city networks \cite{mamatas2023protocol}. 

However, most existing works evaluate the performance of the CP detectors based on the typical trade-off between detection delay in data instances and false alarm rate. Therefore, important aspects related to the computational demands, e.g., resource utilization, actual detection delay and scalability properties, remain under-discussed, with only a few exceptions delving into the theoretical computational complexity \cite{xie2023window} or energy consumption \cite{li2023accurate}. Nevertheless, when dealing with edge devices, computational cost plays a crucial role and may influence the detection efficiency. For instance, a computationally complex algorithm may perform poorly under limited resources, while a less accurate but lightweight algorithm may lead to quicker responses. 

To address this research gap, we propose an experimentation methodology and corresponding facility built on top of real edge cloud test-beds, that pragmatically assesses the online performance of alternative CP procedures. Moreover, we provide support for crucial performance metrics related to resource utilization, i.e., CPU and memory consumption, and detection delay (DD), including actual DD and average response time. Note that our experimentation facility is fully configurable and extensible, e.g., it can be expanded with additional CP mechanisms. For example, it seamlessly incorporates a containerized implementation of MATLAB functions within our novel cloud-native AD evaluation framework. 

In this context, we investigate the applicability of two major CP detection approaches, i.e., Bayesian and cumulative sum (CUSUM), in edge environments. Also, we assess their various performance trade-offs in terms of actual DD (i.e., experimentally measured) and DD in data instances (i.e., theoretical), as well as edge cloud resource utilization. Lastly, we examine their behavior across two test-beds with different configurations to ensure that our insights do not depend on server hardware.  

The major contributions of this work are enlisted below:
\begin{enumerate}
\item We discuss key aspects of selected CP detectors with respect to their employment in edge environments. 
\item We introduce a novel experimentation methodology and facility for assessing CP procedures, based on kubernetes workflows and containerized MATLAB functions.
\item 
We present valuable insights for the performance of the considered procedures in real-world Kubernetes-based edge cloud deployments.
\end{enumerate}








The remainder of this paper is organized as follows. In Section II, we outline the background concepts and the statistical properties of the CUSUM and Bayesian-based CP detectors. In Section III, we present our experimentation methodology, while in Section IV, we detail our proof-of-concept evaluation over the two experimentation test-beds. Finally, in Section V, we provide our concluding remarks.

\section{Considered Detection Mechanisms} \label{detection_mechs}

Here, we give the theoretical background of this work, reviewing main CP detectors in the edge cloud context. Sequential CP analysis \cite{basseville1993detection} refers to the problem of identifying abrupt changes in the underlying characteristics of a time-series as soon as possible, i.e., in an online fashion, with a fixed false alarm rate. According to the particular assumptions for the given time-series data-structure, CP procedures are further categorized to parametric and non-parametric approaches \cite{polunchenko2012state}. Moreover, considering the prior knowledge of the CP occurrence distribution, Bayesian procedures treat the time of a CP as a random variable with a known prior, whereas, non-Bayesian approaches assume a fixed CP without a prior of its location \cite{poor2009quickest}. Next, we discuss in brief the statistical properties of the two aforementioned CP branches.

Regarding the non-Bayesian framework, CUSUM detectors rely on a straightforward and computationally lightweight implementation, since they are based on the calculation of a sequential formula that exclusively involves the current data sample and the previous values of the test statistic \cite{xie2023window}. Additionally, under a fixed false alarm rate, they are known to achieve minimum detection delay \cite{moustakides1986optimal}. CUSUM detectors are non-parametric by design, but can also be applied in a parametric setup, as discussed in \cite{aue2013structural}. However, this application introduces increased computational costs due to the utilization of likelihood estimators, e.g., maximum likelihood estimator (MLE), for the computation of model parameters.

Turning to Bayesian procedures, these  are typically parametric and are based on the fundamental notion of the Bayesian inference. In this context, the seminal work in \cite{adams2007bayesian}, named as Bayesian on-line change point detector (BOCPD), applies a recursive message-passing algorithm to model the posterior distribution $p(r_t|y_t)$ over the run-length $r_t$, i.e., the elapsed time since the last observed CP, given the observed data $y_t$. In contrast to the CUSUM detectors, Bayesian framework derives: i) the quantification of uncertainty in both the number and the location of the CPs \cite{ruggieri2016exact}, and, ii) improved accuracy for short data samples, since the threshold does not rely on the asymptotical behavior of the test statistic\cite{chopin2007dynamic}. However, these benefits come with the cost of strong assumptions for the data-structure, and, an increased computational complexity due to the computation of the posterior distribution in each data instance, with respect to the run-length \cite{figliolia2020fpga}.   

In summary, when considering their general properties, it should be noted that CUSUM procedures can offer advantages in edge cloud systems, including: i) \textit{handling unknown underlying data distributions}, especially in a non-parametric setup, due to their model-free nature; ii) \textit{demonstrating resource efficiency} as they are computationally simple, since they practically compare a fixed pre-change window with the incoming data sequence; iii) \textit{accommodating large amounts of data}, since their threshold is derived asymptotically. On the other hand, Bayesian approaches offer the following benefits: i) \textit{improved performance} in data streams with known distributions; and ii) \textit{stable detection accuracy} for both short and larger data samples. However, these advantages come with the cost of being computational intensive procedures, which may be an issue in resource-constrained edge deployments. 

In the following subsections, we provide a brief overview of the evaluated online CP methods. Note that the presented procedures are being configured with algorithmic extensions to better align with the demands of edge cloud systems.

\subsection{CUSUM-based}
Both CUSUM procedures are implemented based on our novel algorithmic framework detailed in \cite{skaperas2020real} and \cite{skaperas2019real}, rather than being applied directly as open sequential procedures. In detail, we employ a window-based sequential procedure, restricted to $l$ data instances, while we also include a pre-processing step, based on the off-line CP detector given in \cite{aue2013structural}, i.e., for the determination of the training sample $m$. This framework also allows for multiple CPs detection, since it is being executed iteratively. Note that, this modified approach incurs an additional processing cost at the end of each period of length $l$, due to the re-computation of the off-line CP detector in order to calculate the new training period.

Then, we employ a window-based sequential CUSUM approach relying on the following stopping rule $\tau(m)$,
\begin{equation}\label{stoping_time}
\tau(m)=\min\lbrace{t\in{\mathbb{N}:T_{s}(m,t)\geq{C}_{\alpha,g}} \rbrace},
\end{equation}
where a CP is detected in case the test statistic $T_s(m,l)$ in the time window $t\in[1,l]$ exceeds a threshold function $C_{a,\gamma}$. The boundary is defined as $C_{a,\gamma}=\text{cv}_{\alpha}g_{\gamma}$, where $\text{cv}_{\alpha}$ is a critical value calculated from the asymptotic convergence of the fraction $T_s(m,l)/g_{\gamma}$ and $\gamma\in[0,0.5)$ is the tuning parameter of the weight $g_{\gamma}$.

\subsubsection{Non-parametric CUSUM} We follow the work in \cite{fremdt2014asymptotic}, in which the statistical assumptions for the time-series $y_t$ allows even (weak) non-linear dependencies. The CUSUM detector is given by,
\begin{equation}
T_s = \frac{1}{\widehat{\omega}_m}\left(\sum_{j=m+1}^{m+l} y_j - \frac{1}{m}\sum_{j=1}^{m}{y_j} \right),  
\end{equation}
where $\widehat{\omega}^{2}_{m}$ is a suitable estimator of the long-run variance $\omega_{m}^2=\sum_{i\in\mathbb{Z}}\text{cov}(y_0,y_k)$, $\text{cov(.)}$ symbolizes the autocovariance. Here, $\widehat{\omega}_{m}$ captures the standard error under serial dependence and it is approached using the Bartlett kernel estimator,
$\widehat{\omega}_{m}^{2} = \widehat{\kappa}_0 + 2\sum_{j=1}^{W-1}k(\frac{j}{W})\kappa_j
$, where $W$ the window length, $\kappa(.)$ the empirical autocovariance, and, $k(.)$ the Bartlett kernel.

Next, the threshold $C_{\alpha,\gamma}$ applies the weight function,
\begin{equation}
g_{\gamma}=\sqrt{m}\left(1+\frac{l}{m}\right)\left(\frac{l}{m+l}\right)^{\gamma}, \gamma\in[0,\frac{1}{2})
\end{equation}
and the critical value,
\begin{equation}
\text{cv}_{\alpha}= \sup_{t\in\left[0,1\right]}{\frac{W(t)}{t^\gamma}}, 
\end{equation}
$W(t)$ denotes the standard Brownian motion and is approximated using Monte Carlo simulations of its paths on a grid.

It is important to note that the computational cost of the sequential procedure depends exclusively on the computation of the CUSUM detector defined in eq. (2). The boundary function $C_{\alpha,\gamma}$, consisting of eq. (3) and eq. (4), is predefined, and its computation is independent of the provided data.  

\subsubsection{Parametric CUSUM} The CP procedure \cite{aue2015reaction} implements a CUSUM of residuals detector based on an ARMA fit, which is advantageous for linear data structures. Here, the parameters of the ARMA$(p,q)$ model, described by the vector $\widehat{\beta}=(\mu_0,p_0,q_0,\sigma_{0}^{2})$, are estimated in the training period $m$ and the residuals of the process are given as below,
\begin{equation}
\label{arma_res}
    \hat{\epsilon}_{t}=\hat{y}_{t}-\sum_{j=1}^{p}{\hat{p}_{j0}}{\hat{y}_{t-j}}-\sum_{j=1}^{q}{\hat{q}_{j0}}{\hat{\epsilon}_{t-j}},
\end{equation}
Then, the detector is equal to that in eq. (2), replacing the raw data $y_t$ with the residuals $\hat{\epsilon}_t$. As previously, the boundary function $C_{\alpha,\gamma}$ is pre-calculated and does not influence the computational complexity. On the other hand, the estimation of the vector $\hat{\beta}$ of the parameters of the ARMA model, essentially increases the computational cost of the procedure, compared to the non-parametric CUSUM.


\subsection{Bayesian-based}
BOCPD applies the following message-passing algorithm to recursively estimate the run-length distribution $p(r_t|y_{1:t})$,
\begin{equation}
p(r_t|y_{1:t}) \propto	\sum_{r_{t-1}}p(r_t|r_{t-1})p(y_t|r_{t-1},y_{1:{t-1}})p(r_{t-1}|y_{1:t-1}),
\end{equation}
which involves the conditional prior of the run-length $p(r_t|r_{t-1})$, the predictive probability of the model $p(y_t|r_{t-1},y_{t-1})$ and the previous iteration of the recursion. In \cite{alami2020restarted}, the \textit{restarted Bayesian online change-point detector} (r-BOCPD) is introduced as a pruned version of the BOCPD. Assuming that $y_t$ follows the Bernoulli distribution, the change criterion of the r-BOCPD is given by the formula,
\begin{equation}
R_{r:t} = \mathbb{I}\lbrace{ \exists{s}\in(r,t]: \vartheta_{r,s,t} > \vartheta_{r,r,t} \rbrace},
\end{equation}
where the posterior $\vartheta_{r,s,t}=p(r_t=t-s|y_{s:t})$ is the weight assigned to the forecaster $s$ at time $t$ for the starting time $r$, estimated with the recursive formula (initially $\vartheta_{r,1,1} =1$),   
\begin{equation}
\vartheta_{r,s,t} = \begin{cases}
\frac{\eta_{r,s,t}}{\eta_{r,s,t-1}}\text{exp}(-l_{s,t}\vartheta_{r,s,t-1}), & \forall{s}<t \\
\eta_{r,t,t} \times \mathcal{V}_{r,t-1}, & s=t,
\end{cases}
\end{equation}
the Hazard $\eta_{r,s,t}\in(0,1)$ is the hyper-parameter of the algorithm (that controls the trade-off between false alarm and detection delay) and $\mathcal{V}_{r:{t-1}}=\text{exp}\left( -\widehat{L}_{r:{t-1}} \right)$ the initial weight, $\widehat{L}_{r:t-1}=\sum_{s'=r}^{t-1}l_{s':{t-1}}$ is the cumulative cumulative loss incurred by the forecaster $r$ in the interval $[s.t-1]$. Analogously, $l_{s,t} = -\text{log}Lp(y_t|y_{s:{t-1}})$ is the instantaneous loss of the forecaster $s$ at time $t$, and,
\begin{equation}
Lp(y_{t+1}|y_{s:t}) = \begin{cases}
\frac{\sum_{i=s}^{t}y_{i}+1}{t-s+2} & \text{if } y_{t+1} = 1 \\
\frac{\sum_{i=s}^{t}(1-y_i)+1}{t-s+2} & \text{if } y_{t+1} =0,
\end{cases}
\end{equation}
is the Laplace predictor.

\begin{figure}
  \centering
\includegraphics[width=0.87\columnwidth]{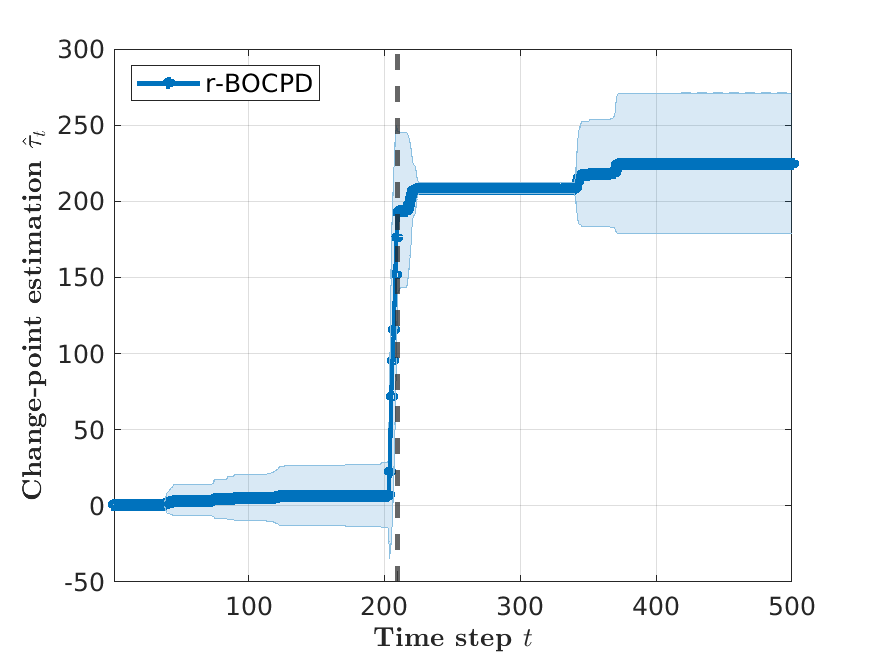}
  \caption {r-BOCPD operation
  , along with the error bars (shadow blue) and the modified stopping rule (dashed dark line).}
  \label{fig:scenario2}
\end{figure}

r-BOCPD calculates $\vartheta_{r,s,i}$ of each $s\in(r,i]$, $i=1,2,\dots,t$. If $R_{r,i}=1$ a CP is detected, the algorithm restarts a new forecaster at $r=t+1$ and deletes the forecasters created before time $t$. Regarding to the computational cost of the procedure, two key factors have to be considered: 1) the chosen starting time $r$, and, 2) the number of algorithm's re-runs $n$, which are executed to achieve a convergence for the probabilistic framework. 

In Fig. 1, we illustrate the operation of r-BOCPD across a piece-wise Gaussian process with a CP at time instance $t=200$, averaged over $n=500$ runs, where the error bars before and after $t=200$ denote false alarms. As shown, r-BOCPD provides a descriptive solution for the estimation of the CPs, which increases its explainability, but also poses challenges for its applicability, since it does not derive fixed CP detections; necessary in automated decision making algorithms. For this reason, we modified the Bayesian detector to handle the false alarm rate and derive fixed CP detections $\hat{\tau}_t$, through the application of following stopping rule,
\begin{equation}
\hat{\tau}_t = \min\{t\in\mathbb{N}: \overline{cp}_t > tq \}, 
\end{equation}
where $\overline{cp}_t$ symbolizes the CP detections of r-BOCPD at time instance $t$, averaged over $n$ number of runs and $q$ is a weight parameter that serve to handle the false alarm rate. Regarding Fig. 1 outcomes, the fixed CP detection using the modified stopping rule is depicted with the dashed dark line.  



\section{Evaluation Methodology}


Our experimentation methodology assesses recent, representative AD mechanisms in real edge cloud environments, attempting to uncover the interplay between theoretical properties of algorithms and their impact on edge cloud resources.

We consider cloud-native AD workflows implemented as Kubernetes argo workflows\footnote{https://github.com/argoproj/argo-workflows}. We assume that new AD workflows are being deployed, e.g., somethings happens and raises the criticality level as high, operate for a time period or transmit a fixed number of samples, and then the workflows are being removed. The workflows utilize AD clients and servers with respect to given configurations. 

The clients are able to reproduce traces of measurements or generate anomalies by embedding an arbitrary number of time-series with indicative structures. At this point of investigation, we focus on the impact of AD workflows on resource utilization, rather than on the theoretical accuracy of algorithms. The latter is extensively studied in other works, e.g., \cite{skaperas2019real}.  
The servers are MATLAB containers supporting all considered mechanisms, enabled and configured on-demand. The communication is based on a custom REST API. The AD server Pods also host side-car containers that trace resource utilization of mechanisms, based on an extension of Kubernetes Network Benchmark tool\footnote{https://github.com/InfraBuilder/k8s-bench-suite/tree/master}. 

In practice, clients communicate with a server hosting the CP detectors in an on-line manner and transmit each new data sample iteratively (every 100 ms). Subsequently, the server processes the incoming data using the selected CP mechanism, and responds if an anomaly is detected. This setup allows us to specify an experiment configuration with the details of AD workflows (e.g., number of clients, cluster node affinity, etc.), mechanisms (e.g., type and configuration), as well as of communicated and processed data (e.g., out of a trace-file or generated from known distributions/time-series models).

The performance metrics include: (i) the DD in data instances: number of data points $t$ needed to be processed for the detection of a CP after its occurrence,
(ii) the actual DD: time duration between the CP occurance and its estimation (measured in ms), including the data communication and the processing,
(iii) the CPU and memory consumption: combined impact of AD workflows, i.e., including pod deployment/removal, data communication / storage / processing, etc, in percentage (\%), and, (iv) the response time: the total time for a data point to be transmitted from the client, processed by the server's containerized AD mechanism, and for the server to send a response back to the client, measured in ms.

The experiments were replicated in two separate test-beds to also account for the impact of hardware on results. The first test-bed, at the University of Macedonia (UOM), has 2 Dell PowerEdge R630 servers with dual Intel(R) Xeon(R) E5-2620 v4 @ 2.10GHz 16-core CPUs and 50GB RAM. The second test-bed, at ATHENA Research Center, features a Dell PowerEdge T640 server with an Intel(R) Xeon(R) Silver 4210R @ 2.40GHz 16-core CPU and 64GB RAM. 

The real operation of AD mechanisms is hosted in Kubernetes clusters of one master and two worker nodes. The three nodes are VMs with limited resource availability (e.g., 2 CPU cores and 4GB of RAM) to resemble resource-constrained edge environments. In the UOM test-bed, the VMs are distributed across both servers, accounting for physical network connectivity considerations. All VM nodes are running Ubuntu 22.04.2 LTS (with kernel version 5.15.0-71-generic). The kubernetes cluster is being built based on kubeadm (v1.28.2), with kubectl and kubelet also aligned with the same version and containerd maintained at version 1.6.24. The Argo workflow is set to v3.4.4. Furthermore, both test-beds use the XCP-ng\footnote{https://xcp-ng.org} virtualization platform and the Flannel (v0.22.3)  container networking plugin\footnote{https://github.com/flannel-io/flannel} for the nodes' and pods' inter-communication.


\begin{figure*}[t]
\centering
    \begin{subfigure}[b]{0.29\textwidth}       \includegraphics[width=\textwidth]{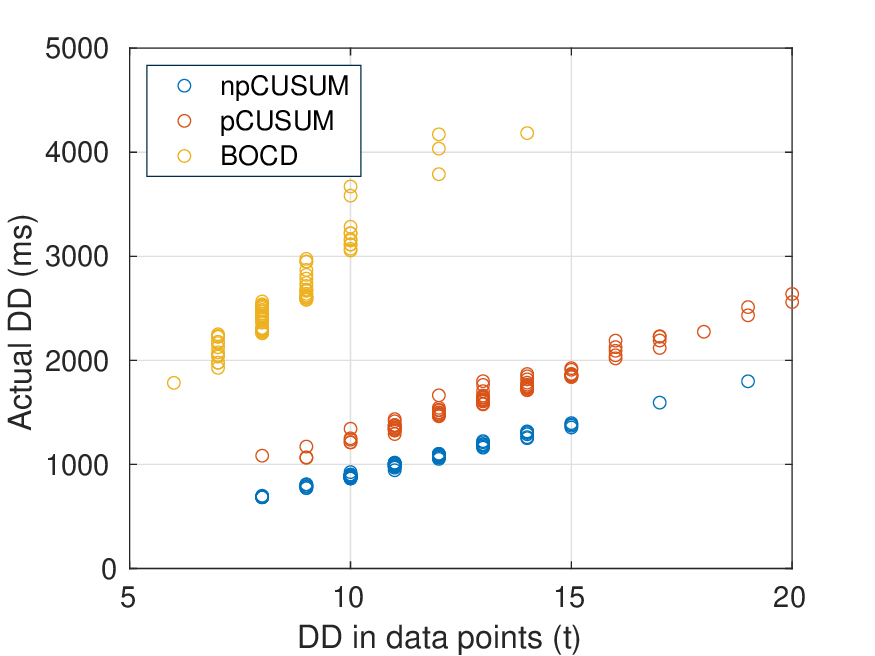}
    \caption{actual DD vs DD in data points.}
    \label{fig:client_11a}
    \end{subfigure}%
    \begin{subfigure}[b]{0.29\textwidth}       \includegraphics[width=\textwidth]{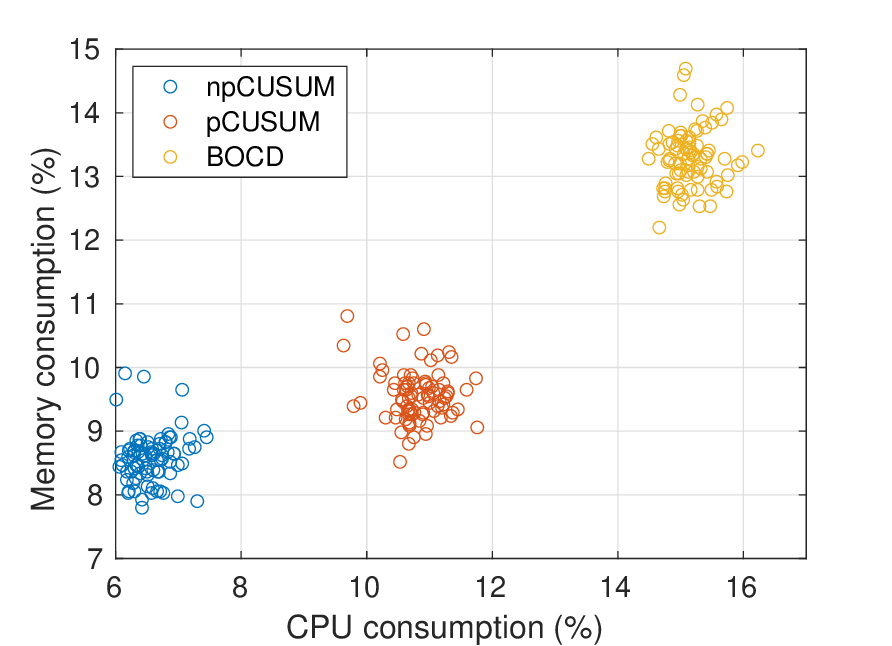}
    \caption{Memory vs CPU consumption.}
    \label{fig:client_11b}
    \end{subfigure}%
    \begin{subfigure}[b]{0.29\textwidth}   \includegraphics[width=\textwidth]{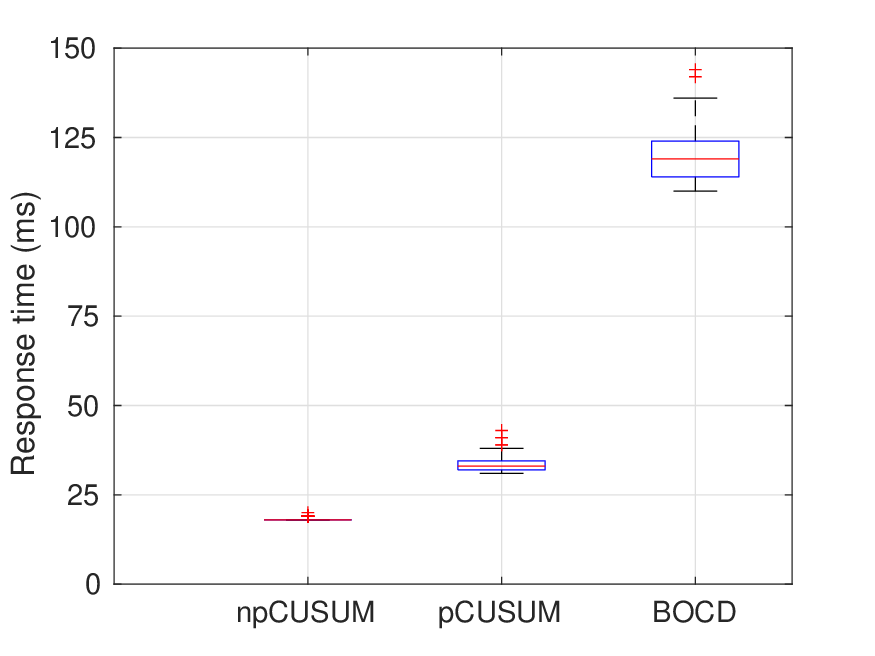}
    \caption{Response time for each data point.}
    \label{fig:client_11c}
    \end{subfigure}
    
    \begin{subfigure}[b]{0.29\textwidth}       \includegraphics[width=\textwidth]{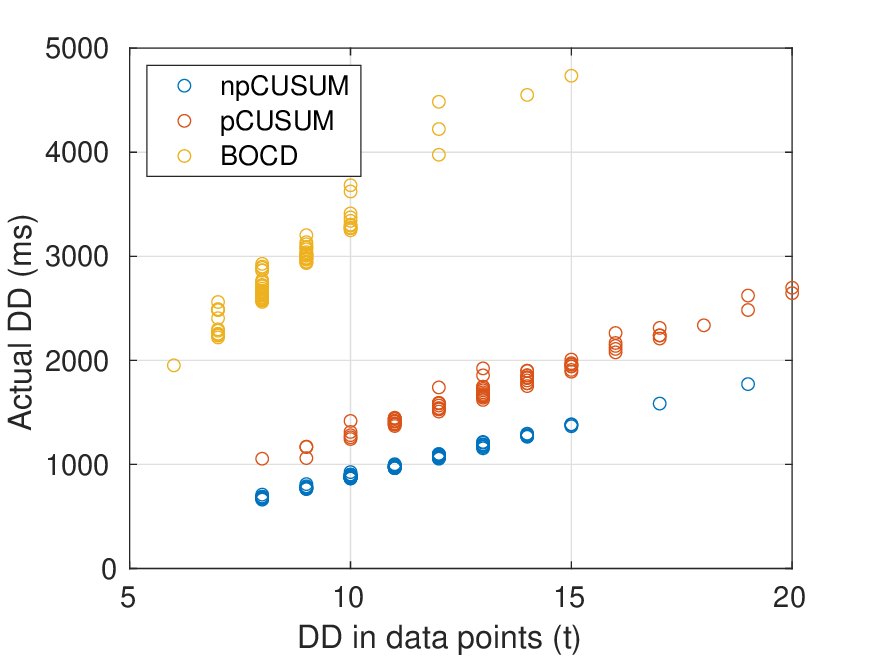}
    \caption{actual DD vs DD in data points.}
    \label{fig:client_11d}
    \end{subfigure}%
    \begin{subfigure}[b]{0.29\textwidth}      \includegraphics[width=\textwidth]{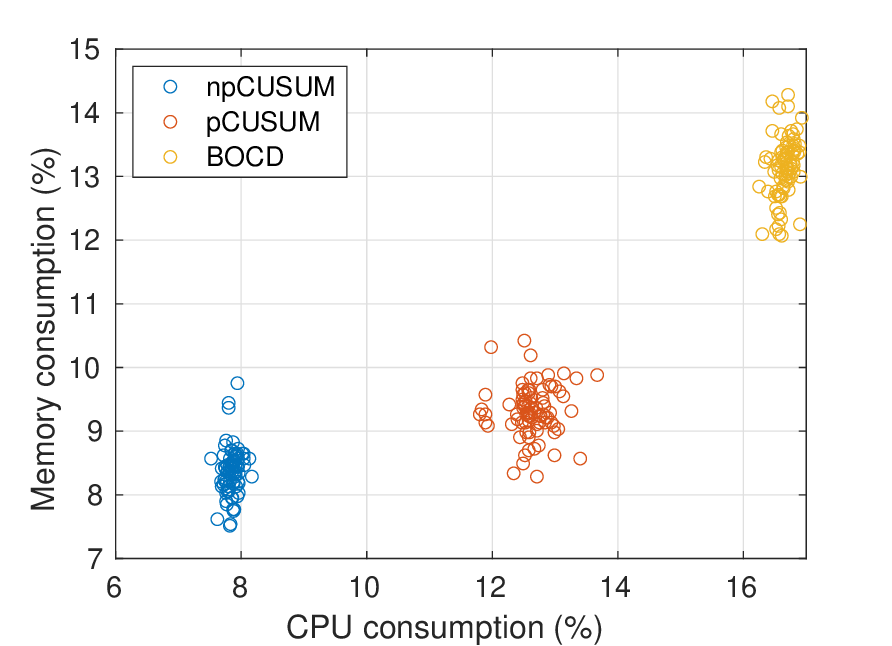}
    \caption{Memory v CPU consumption.}
    \label{fig:client_11e}
    \end{subfigure}%
    \begin{subfigure}[b]{0.29\textwidth}   \includegraphics[width=\textwidth]{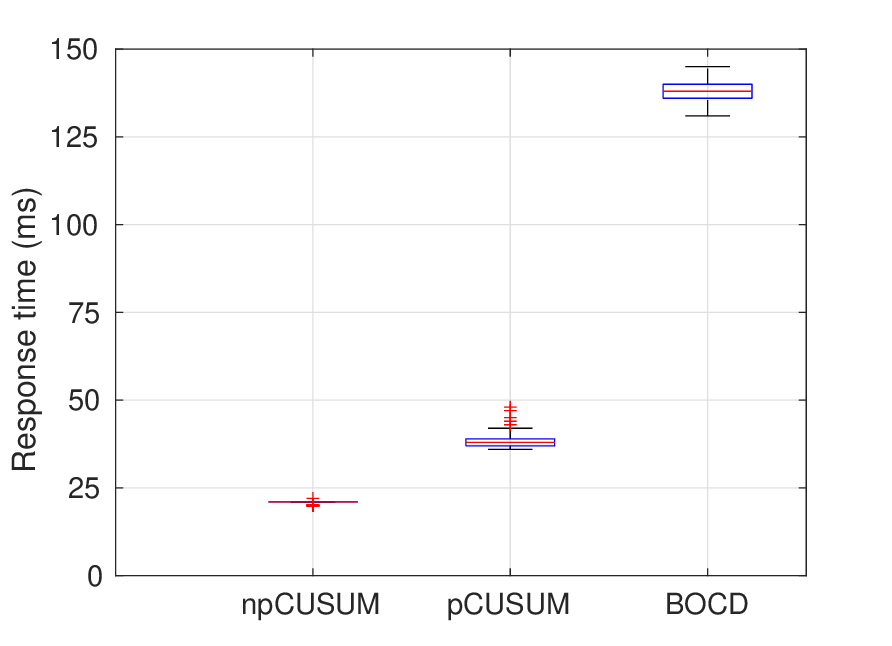}
    \caption{Response time for each data point.}
    \label{fig:client_11f}
    \end{subfigure}
    \caption{i) actual DD versus DD in data points, ii) memory versus CPU consumption, and, iii) response time for each time-series being processed on-line, assuming 1 client. Applying npCUSUM, pCUSUM and BOCD methods, in both UoM test-bed (first row) and ATHENA test-bed (second row).} 
    \label{fig:client_1}
\end{figure*}


\section{Experimental Results} 
In this section, we primarily focus on the associated performance issues of the adaptation of the CP procedures in real online AD applications (e.g., measuring the actual response time), rather than statistical aspects, such as the detection accuracy with respect to different types of data structures, which are the main subjects of other studies.

The considered CP methods: i) non-parametric CUSUM (npCUSUM), ii) ARMA-based CUSUM (pCUSUM), and, iii) r-BOCPD (BOCD), are being assessed over synthetic time-series across the UOM (t1) and ATHENA (t2) test-beds (i.e., being exchanged between containarized AD clients and servers implemented in the form of Kubernetes workflows). The parameters of the examined procedures, for a given significance level $\alpha = 0.05$, follows: For CUSUM, the tuning parameter $\gamma=0.25$, the training period $m=100$ and the window size $l=50$. Regarding the r-BOCPD, the tuning parameter $\eta_{r,s,t}=\frac{1}{t-s+1}$, the starting time instance $r=1$, the number of runs $n=100$, and the weight parameter $q=0.95$. The synthetic time-series $X_n$ have been produced based on randomly generated values from an ARMA(1,1) auto-regressive moving average model (as in \cite{skaperas2019real} or \cite{aue2015reaction}), with parameters $\phi=0.4$, $\theta=0.2$ and innovations $\epsilon_t\sim\mathbf{N}(0,1)$. The sample size is $T=500$, while a CP is introduced at $t_{cp}=250$ by shifting the mean value from $\mu=0$ to $\mu=1$. Finally, results are assessed over 100 Monte Carlo simulations.

\begin{table}[t]
\centering
\renewcommand{\arraystretch}{0.8}
\setlength{\tabcolsep}{6pt}
\caption{Detection accuracy vs detection delay trade-off.}
\label{tab:t2}
\begin{tabular}{|c|c|c|}
\hline
\multicolumn{1}{|c|}{ } & \multicolumn{1}{|c|}{True (false) alarm rate} & \multicolumn{1}{|c|}{CP gap} \\
\hline
\multicolumn{1}{|c|}{npCUSUM} & \multicolumn{1}{|c|}{0.95 (0.05)} & \multicolumn{1}{|c|}{10} \\
\hline
\multicolumn{1}{|c|}{pCUSUM} & \multicolumn{1}{|c|}{1 (0)} & \multicolumn{1}{|c|}{12} \\
\hline
\multicolumn{1}{|c|}{BOCD} & \multicolumn{1}{|c|}{0.94 (0.06)} & \multicolumn{1}{|c|}{7} \\
\hline
\end{tabular}
\end{table}

    

Initially, we discuss the trade-off between detection accuracy (true/false alarm rate) and detection delay (gap between the actual and estimated CP in data points, $t_{\widehat{cp}}-t_{cp}$). According to Table \ref{tab:t2}, pCUSUM achieves 100\% true alarm detection but at the cost of increased detection delay. On the other hand, BOCD concludes on the shortest delay being more prone to over-estimation error, while npCUSUM provides a compromise between over-estimation and detection delay.


Fig. \ref{fig:client_1} illustrates the metrics: i) actual DD, ii) DD in data points, iii) memory and CPU utilization in percentage (\%); and, iii) the response time (in ms), for all three CP procedures, in t1 (first row) and t2 (second row) test-beds, assuming that 1 client utilizes the service. As shown in Fig. \ref{fig:client_11a}, for t1 test-bed, DD in data points may not suffice to determine the "fastest" procedure. For example, when comparing BOCP with npCUSUM and pCUSUM, we observe that the former, results on a significantly larger actual DD, although it provides the shortest DD in data points. Moreover, we observe a linear relationship between the actual DD and DD measured in data points, with BOCP and pCUSUM displaying greater fluctuations (due to the complexity of the detectors). Analogous results are derived for the t2 test-bed, illustrated in Fig. \ref{fig:client_11d}.     
 
Figs. \ref{fig:client_11b} and \ref{fig:client_11e} depict the CPU and memory consumption of each CP method. Concerning both t1 and t2 test-beds, the npCUSUM method exhibits superior performance in terms of both CPU and memory consumption, followed by the pCUSUM, with the BOCD method exhibiting the higher resource demands. Note that, despite the fluctuations in the CPU utilization between the two test-beds, an almost stable performance is indicated for the examined methods in terms of resource utilization. This implies that the CP detectors exhibit a degree of independency in relation to the physical machine configuration (i.e., the algorithms use the same amount of resources in the virtual space). Consequently, the mechanisms may be characterized by a relatively stable resource footprint, which can be valuable to Kubernetes resource-optimization tasks (e.g., elasticity). However, this issue is complex enough to deserve additional extensive studies.          



Finally, figs. \ref{fig:client_11c} and \ref{fig:client_11f}, illustrate that the response times for each method remain relatively stable in both test-beds, with minimal variation, especially in the case of the npCUSUM method. Note that, the higher variation in response time regarding to the BOCD procedure, reflects to the variation of the actual DD for similar DD in data points. 

\begin{figure}[t]
\centering
    \begin{subfigure}[b]{0.25\textwidth}       \includegraphics[width=\textwidth]{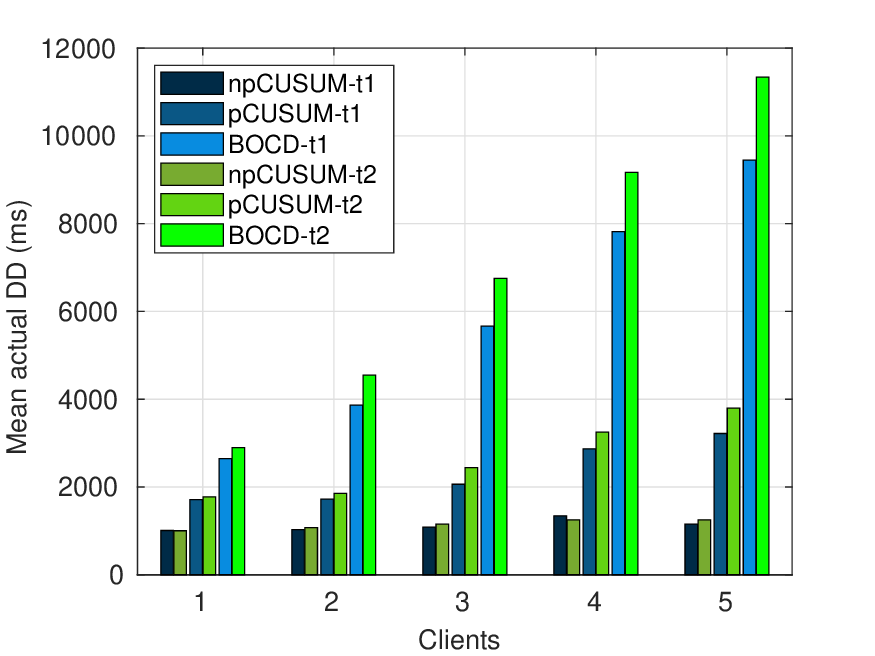}
    \end{subfigure}%
    \begin{subfigure}[b]{0.25\textwidth}   \includegraphics[width=\textwidth]{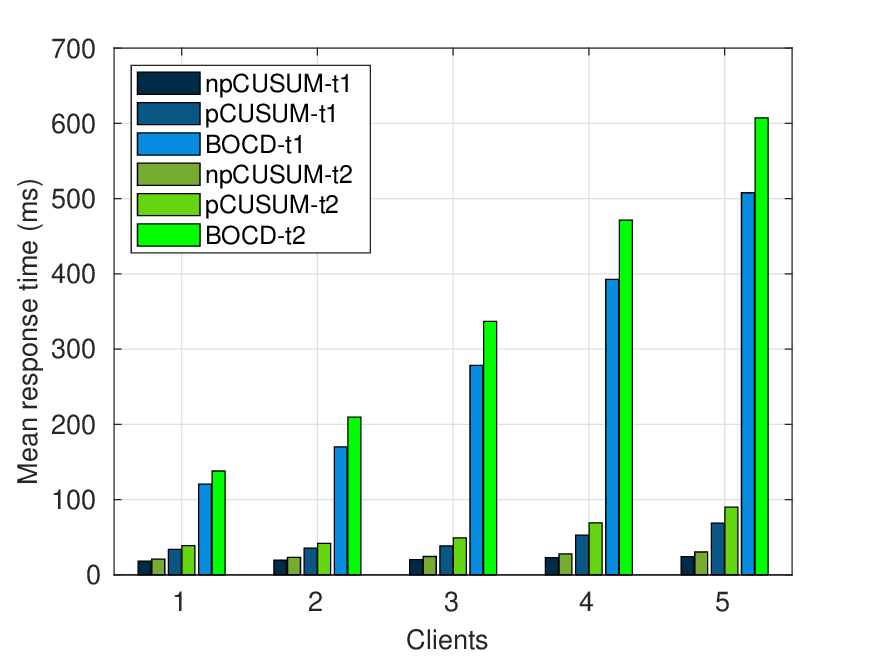}
    \end{subfigure}
    \begin{subfigure}[b]{0.25\textwidth}       \includegraphics[width=\textwidth]{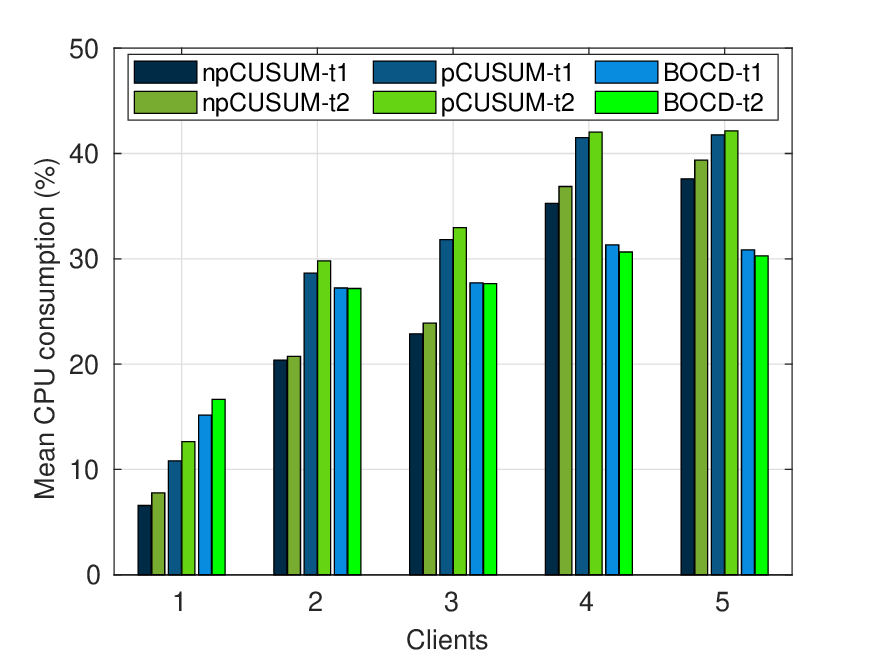}
    \end{subfigure}%
    \begin{subfigure}[b]{0.25\textwidth}   \includegraphics[width=\textwidth]{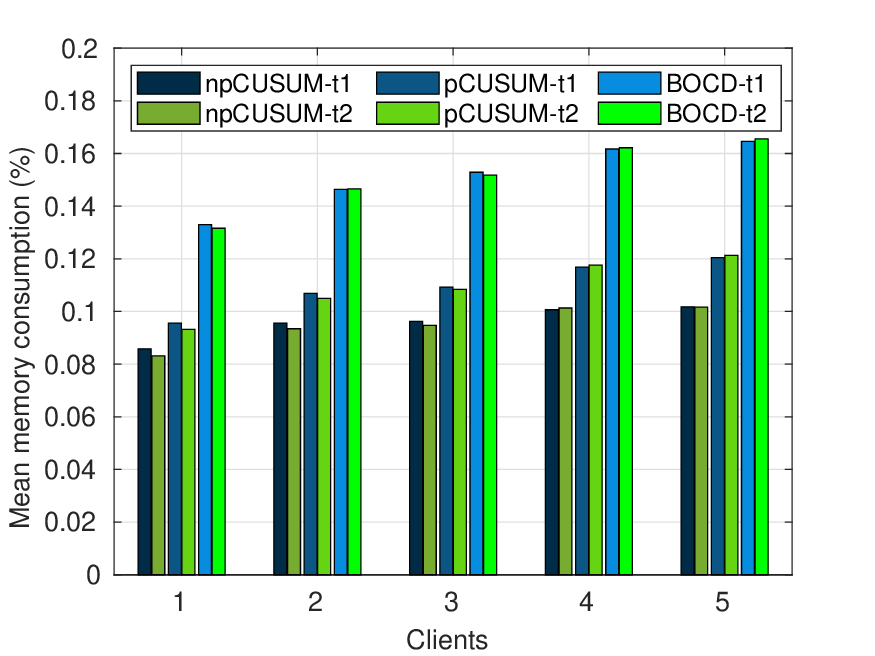}
    \end{subfigure}
    \caption{Mean: i) actual DD, ii) response time, iii) CPU consumption, and, iv) memory consumption, in UoM (t1) and ATHENA (t2) test-beds, regarding $k=\{1,\cdots,5\}$ clients.}
    \label{fig:fig2_all}
\end{figure}

Subsequently, Fig. \ref{fig:fig2_all} illustrates the mean actual DD, response time, and, CPU and memory utilization, for $k=\{1,\dots,5\}$ clients served simultaneously. Concerning the actual DD in Fig. 3a, we observe that an increase in the number of clients mainly affects pCUSUM and BOCD, resulting in a $2$ to $4$ times increase in the actual DD, when compared to the scenario with one client. On the other hand, npCUSUM shows a relative stable performance irrespective of the number of clients to be processed. This result is also confirmed by the corresponding measurements of the response times in Fig. 3b.  

In addition, in Figs. 3c and 3d, we identify an increase in the resource consumption of the CP procedures, which, however, retain distinct characteristics in terms of CPU and memory utilization. This fact may be attributed to the initiation and termination of pods that temporarily impact CPU performance, i.e., this may be significant in the cases with a larger number of clients and a short experiment duration. However, this aspect requires further investigations, e.g., to decouple algorithmic resource consumption from pod manipulation overhead. Nevertheless, BOCD exhibits a superior performance in terms of CPU consumption.  Interestingly, the increase in the resource consumption of npCUSUM does not impact its actual DD, due to its very lightweight algorithm.  

We now enlist the most important gained insights with respect to the actual DD and response times of the selected CP methods. More precisely:
\begin{itemize}
\item npCUSUM provides superior performance and offers scalability advantages, as the number of clients increases.
\item pCUSUM results in an increased actual DD, when compared to npCUSUM, especially when dealing with a larger number of concurrently served clients, which is attributed to the involved MLE calculations.  
\item BOCD exhibits significantly higher actual DD than CUSUM methods, posing scalability challenges for more clients and highlighting the computational intensity of Bayesian procedures.
\item the typical trade-off between false alarm and detection delay is not sufficient to characterize the performance of CP mechanisms in real AD applications, especially of those deployed at the edge. 
\end{itemize}

Lastly, turning our attention to the resource consumption, the three CP detectors provide a discrete resource utilization behavior, with slight variations within each method, which is also maintained across both test-beds. This observation intuitively suggests that resource consumption may exhibit some degree of predictability, which is an important input for resource optimization mechanisms.

\section{Conclusions} 
In this paper, we provided a fresh view on the evaluation of state-of-the-art (CP based) AD algorithms in edge cloud systems. In our comparative analysis, we utilized a novel methodology and a cloud-native system for the online assessment of CP algorithms. This framework allowed us to incorporate additional metrics to characterize the computational complexity and the performance of the examined procedures, such as resource consumption in terms of CPU and memory as well as actual response time. Our initial results demonstrated that, on one hand, the statistical properties of CP algorithms are generally consistent with their behavior in edge cloud implementations, and on the other hand, underlined limitations of the typical evaluation metrics for CP procedures.


\section{Acknowledgments}
This work is partially supported by the Horizon Europe CODECO Project under Grant number 101092696.

\bibliographystyle{IEEEtran}
\bibliography{IEEEabrv,main.bib}

\end{document}